\documentclass[pra,preprint,amsmath,amssymb]{revtex4}
\usepackage{bm}
\usepackage{amsmath,amssymb}
\usepackage{color}
\usepackage{amsthm} 

\theoremstyle{plain}
  \newtheorem{theorem}{Theorem}

\theoremstyle{definition}
  \newtheorem{definition}{Definition}

\theoremstyle{remark}
  
\theoremstyle{plain}
  \newtheorem*{theorem*}{Theorem}
  \newtheorem*{lemma*}{Lemma}
  \newtheorem*{corollary*}{Corollary}
  \newtheorem*{proposition*}{Proposition}
  \newtheorem*{claim*}{Claim}

\begin{document}
\title{Modified group non-membership is in AWPP}
\author{Tomoyuki Morimae}
\affiliation{ASRLD Unit, Gunma University, 1-5-1 Tenjincho, Kiryushi,
Gunma, 376-0052, Japan}
\email{morimae@gunma-u.ac.jp}
\author{Harumichi Nishimura}
\affiliation{Graduate School of Information Science, Nagoya University,
Furocho, Chikusaku, Nagoya, Aichi, 464-8601, Japan}
\email{hnishimura@is.nagoya-u.ac.jp}
\author{Fran\c{c}ois Le Gall}
\affiliation{Department of Computer Science, The University of Tokyo,
7-3-1 Hongo, Bunkyoku, Tokyo, 113-8656, Japan}
\email{legall@is.s.u-tokyo.ac.jp}

\begin{abstract}
It is known that the group non-membership problem is in 
QMA relative to any group oracle
and in ${\rm SPP}\cap{\rm BQP}$ relative to group oracles for
solvable groups.
We consider a modified version of the group non-membership problem
where the order of the group is also given as an additional input.
We show that  
the problem is in AWPP relative to any group oracle.
To show the result, we use the idea of the
postselected quantum computing.
\end{abstract}

\maketitle

\section{Introduction}
The group non-membership (GNM) is the following problem:
\begin{itemize}
\item
Input: Group elements $g_1,g_2,...,g_k$, and $h$ in some finite group $G$.
\item
Question: Is $h\notin H\equiv\langle g_1,...,g_k\rangle$?
\end{itemize}
Here, $H\equiv\langle g_1,...,g_k\rangle$ is the group generated
by $g_1,...,g_k$.
This problem has long been studied for black-box groups~\cite{bbg},
which are finite groups whose elements are encoded as strings of a
given length and whose group operations are performed by a group oracle.
The GNM is known to be hard for classical computing: 
for some group oracle $B$,
GNM is not in ${\rm BPP}^B$~\cite{B1,B2}.
Furthermore, it was also shown that
for some group oracle $B$,
GNM is not in ${\rm NP}^B$~\cite{B1,B2},
and
for some group oracle $B$,
GNM is not in ${\rm MA}^B$~\cite{Watrous}.

Upper bounds of GNM have also been derived.
For example, it was shown that 
GNM is in ${\rm coNP}^B$~\cite{bbg},
${\rm AM}^B$~\cite{B1,B2},
and
${\rm QMA}^B$~\cite{Watrous} for any group oracle $B$.
If we restrict the group to be solvable, upper bounds can be
improved: it was shown that
GNM is in ${\rm BQP}^B$~\cite{Watrous_solvable}
and
${\rm SPP}^B$~\cite{Vin_SPP}.

In this paper,
to deepen our understanding of the upper bounds of GNM,
we consider a slightly modified
version of GNM, which we call the modified GNM:
\begin{itemize}
\item
Input: Group elements $g_1,g_2,...,g_k$, and $h$ in some finite group $G$, and 
$|\langle g_1,...,g_k\rangle|$.
\item
Question: Is $h\notin H\equiv\langle g_1,...,g_k\rangle$?
\end{itemize}
In other words, in the modified GNM,
the order $|\langle g_1,...,g_k\rangle|$
of the generated group $\langle g_1,...,g_k\rangle$ is also given
as an additional input.

We show that the modified GNM is in 
AWPP relative to any group oracle.
The class AWPP was introduced by Fenner, Fortnow, Kurtz, and Li~\cite{toolkit}
to understand the structure of counting complexity classes
(see also Refs.~\cite{Li,Fenner}). 
AWPP is also famous among quantum information scientists,
since it is one of the two best upper bounds of BQP~\cite{FR}. 
(The other one is QMA (or QCMA). 
No direct relation is known between QMA and AWPP.
It is at least known that they share the same upper bound, SBQP,
namely,
${\rm QMA}\subseteq{\rm SBQP}$~\cite{Vyalyi} 
and ${\rm AWPP}\subseteq{\rm  SBQP}$.)
Therefore, our result implies
that if GNM is changed to a bit easier problem
by adding an extra input, its upper bound is improved to
the intersection of QMA and AWPP.

The definition of AWPP is as follows.
(Here, we take a simpler definition of AWPP by Fenner~\cite{Fenner}.)
\begin{definition}
\label{def:AWPP}
A language $L$ is in ${\rm AWPP}$ iff 
there exist $f\in {\rm FP}$ and $g\in {\rm GapP}$ such that
for all $w$, $f(w)>0$ and
\begin{itemize}
\item[1.]
If $w\in L$ then $\frac{2}{3}\le \frac{g(w)}{f(w)}\le 1$,
\item[2.]
If $w\notin L$ then $0\le \frac{g(w)}{f(w)}\le \frac{1}{3}$.
\end{itemize}
Here, ${\rm FP}$ is the class of functions 
from bit strings to integers that are computable in polynomial time by a Turing machine.
A ${\rm GapP}$ function~\cite{GapP}
is a function from bit strings to integers that is equal to
the number of accepting paths minus that of rejecting paths of a nondeterministic
Turing machine which takes the bit strings as input.
The FP function $f$ can be replaced with $2^{q(|w|)}$ for a polynomial 
$q$~\cite{GapP,Li},
and the error bound $(\frac{1}{3},\frac{2}{3})$
can be replaced with $(2^{-r(|w|)},1-2^{-r(|w|)})$ 
for any polynomial $r$~\cite{toolkit,Li}. 
\end{definition}

Our proof is based on the idea of postselected quantum computing.
The postselection is a fictious ability that one can always
obtain a specific measurement result even if its occurring
probability is exponentially small.
The class of languages that can be efficiently recognized
by a quantum computer with the postselection is called postBQP,
and it is known that ${\rm postBQP}={\rm PP}$~\cite{postBQP}.
If we consider a restricted version of postBQP
where the postselection probability is close to an FP function 
divided by $2^{poly}$, the class was shown to be equal to
AWPP~\cite{MorimaeNishimura}.
Our proof is based on this relation between postselected quantum
computing and AWPP: we first propose a postBQP algorithm
that can solve the modified GNM, and then show that the postselection
probability satisfies the condition.
Then, by using the relation between the output probability
distribution of quantum computing and GapP function~\cite{FR},
we conclude that the modified GNM is in AWPP.
Our quantum algorithm is based on that of Watrous~\cite{Watrous}.
He showed that if the state $\sum_{g\in H}|g\rangle$, 
which is believed to be hard
to generate with a polynomial-size quantum computing,
is given as a witness, GNM is verified efficiently.
In our algorithm, the witness is generated by polynomial-size
quantum computing with postselection. This result itself means
${\rm GNM}\in{\rm postBQP}={\rm PP}$, which is trivial
since it is already known that 
${\rm QMA}\subseteq{\rm SBQP}\subseteq{\rm PP}$. 
Our contribution is that we 
point out that the postselection probability satisfies
a nice condition, and therefore if the GNM
is modified as described above, it is in AWPP.

\section{Proof}
Now we show our result
that the modified GNM is in AWPP relative to any group oracle.
First, let us remember the group oracle
and a theorem shown by Babai~\cite{B1}.
A group oracle $B$ can be represented by a family of bijections $\{B_n\}$
with each member having the form $B_n:\{0,1\}^{2n+2}\to\{0,1\}^{2n+2}$
and satisfying certain constraints that specify its 
operation (see Section 2 in Ref.~\cite{Watrous} for the
precise definition).
We denote the group associated with each $B_n$ by $G(B_n)$.
In other words, elements of $G(B_n)$ form some subset of $\{0,1\}^n$
and the group structure of $G(B_n)$ is determined by the function $B_n$.
The following theorem by Babai~\cite{B1} (see also Ref.~\cite{Watrous}) is a basis of our result.
\begin{theorem}
\label{theorem:Babai}
For any group oracle $B=\{B_n\}$, 
there exists a randomized procedure ${\mathcal P}$ acting as follows:
On input $g_1,...,g_k\in G(B_n)$ and $\epsilon>0$, the procedure outputs an element of
$H\equiv\langle g_1,...,g_k\rangle$ in time polynomial in 
$n+\log \frac{1}{\epsilon}$ such that each $g\in H$ is output with probability in
the range $(\frac{1}{|H|}-\epsilon,\frac{1}{|H|}+\epsilon)$.
\end{theorem}

As is explained in Ref.~\cite{Watrous}, we can simulate the classical randomized
procedure ${\mathcal P}$ in ``quantum way":
Let us assume that a random bit is generated $s(n)$ times during ${\mathcal P}$,
where $s$ is a polynomial.
We first generate the state 
\begin{eqnarray*}
	\frac{1}{\sqrt{N}}\sum_{z\in\{0,1\}^{s(n)}}|z\rangle,
\end{eqnarray*}
where $N\equiv 2^{s(n)}$, $|z\rangle$ is an $s(n)$-qubit state, and 
each $z$ is an $s(n)$-bit string representing random numbers
generated during ${\mathcal P}$.
By coupling sufficiently many ancilla qubits and running ${\mathcal P}$ for each 
branch controlled by $z$, we obtain
\begin{eqnarray*}
|\Psi\rangle&\equiv&
\frac{1}{\sqrt{N}}\sum_{z\in\{0,1\}^{s(n)}}
|\eta_z\rangle\otimes|\phi_z\rangle\\
&=&\frac{1}{\sqrt{N}}\sum_{g\in H}\sqrt{\gamma_g}|g\rangle\otimes|garbage(g)\rangle,
\end{eqnarray*}
where $\eta_z$ is an element of $H$,
$\phi_z$ is a $t(n)$-bit string 
corresponding to the leftover of the procedure ($t$ is a polynomial), and
\begin{eqnarray*}
|garbage(g)\rangle\equiv\frac{1}{\sqrt{\gamma_g}}
\sum_{z:\eta_z=g}|\phi_z\rangle.
\end{eqnarray*}
Here, $\gamma_g$ is the normalization factor, i.e., 
the number of $z$ such that $\eta_z=g$.
From Theorem~\ref{theorem:Babai},
\begin{eqnarray}
\frac{\gamma_g}{N}\in\Big(\frac{1}{|H|}-\epsilon,\frac{1}{|H|}+\epsilon\Big).
\label{eq0}
\end{eqnarray}
Furthermore, due to the normalization of $|\Psi\rangle$,
\begin{eqnarray*}
\sum_{g\in H}\frac{\gamma_g}{N}=1.
\end{eqnarray*}
Therefore if we write
\begin{eqnarray*}
\frac{\gamma_g}{N}=\frac{1}{|H|}+\epsilon_g,
\end{eqnarray*}
where $-\epsilon\le \epsilon_g\le\epsilon$,
we obtain
$
\sum_{g\in H}\epsilon_g=0.
$
Hence,
\begin{eqnarray}
\sum_{g\in H}\Big(\frac{\gamma_g}{N}\Big)^2
&=&\frac{1}{|H|}+\frac{2}{|H|}\sum_{g\in H}\epsilon_g
+\sum_{g\in H}\epsilon_g^2\nonumber\\
&=&\frac{1}{|H|}+\sum_{g\in H}\epsilon_g^2\label{eq1}\\
&\ge& \frac{1}{|H|}\label{eq1.1}.
\end{eqnarray}
Let us couple $|\Psi\rangle$ with 
$|+\rangle\equiv(|0\rangle+|1\rangle)/\sqrt{2}$ and apply
a controlled multiplication by the element $h$ to obtain
\begin{eqnarray*}
|\Psi'\rangle\equiv\frac{1}{\sqrt{2N}}\sum_{g\in H}\sqrt{\gamma_g}
\Big(|g\rangle\otimes|0\rangle
+|gh\rangle\otimes|1\rangle\Big)\otimes|garbage(g)\rangle,
\end{eqnarray*}
where the second register is the coupled qubit.
Let us apply Hadamard on the second register:
\begin{eqnarray*}
\frac{1}{2\sqrt{N}}\sum_{g\in H}
\sqrt{\gamma}_g\Big[
(|g\rangle+|gh\rangle)|0\rangle+
(|g\rangle-|gh\rangle)|1\rangle
\Big]|garbage(g)\rangle.
\end{eqnarray*}
Let us prepare two copies of them, and add an 
ancilla qubit $|1\rangle_a$:
\begin{eqnarray*}
\frac{1}{4N}\sum_{g,g'\in H}
\sqrt{\gamma}_g\sqrt{\gamma_{g'}}\Big[
&&(|g\rangle+|gh\rangle)(|g'\rangle+|g'h\rangle)|00\rangle|1\rangle_a\\
&+&(|g\rangle+|gh\rangle)(|g'\rangle-|g'h\rangle)|01\rangle|1\rangle_a\\
&+&(|g\rangle-|gh\rangle)(|g'\rangle+|g'h\rangle)|10\rangle|1\rangle_a\\
&+&(|g\rangle-|gh\rangle)(|g'\rangle-|g'h\rangle)|11\rangle|1\rangle_a
\Big]|garbage(g)\rangle|garbage(g')\rangle.
\end{eqnarray*}
Flip the ancilla qubit if the third register of the above state
is in the state $|00\rangle$:
\begin{eqnarray*}
\frac{1}{4N}\sum_{g,g'\in H}
\sqrt{\gamma}_g\sqrt{\gamma_{g'}}\Big[
&&(|g\rangle+|gh\rangle)(|g'\rangle+|g'h\rangle)|00\rangle|0\rangle_a\\
&+&(|g\rangle+|gh\rangle)(|g'\rangle-|g'h\rangle)|01\rangle|1\rangle_a\\
&+&(|g\rangle-|gh\rangle)(|g'\rangle+|g'h\rangle)|10\rangle|1\rangle_a\\
&+&(|g\rangle-|gh\rangle)(|g'\rangle-|g'h\rangle)|11\rangle|1\rangle_a
\Big]|garbage(g)\rangle|garbage(g')\rangle.
\end{eqnarray*}
Note that
\begin{eqnarray*}
\langle+^{\otimes t(n)}|garbage(g)\rangle&=&\frac{1}{\sqrt{\gamma_g}}
\sum_{z:\eta_z=g}\langle+^{\otimes t(n)}|\phi_z\rangle\\
&=&\frac{1}{\sqrt{\gamma_g2^{t(n)}}}\gamma_g\\
&=&\frac{\sqrt{\gamma_g}}{\sqrt{2^{t(n)}}}.
\end{eqnarray*}
Therefore,
if we postselect garbage registers onto $|+\rangle^{\otimes 2t(n)}$,
the (unnormalized) state after the postselection is
\begin{eqnarray*}
\frac{1}{4N2^{t(n)}}\sum_{g,g'\in H}
\gamma_g\gamma_{g'}\Big[
&&(|g\rangle+|gh\rangle)(|g'\rangle+|g'h\rangle)|00\rangle|0\rangle_a\\
&+&(|g\rangle+|gh\rangle)(|g'\rangle-|g'h\rangle)|01\rangle|1\rangle_a\\
&+&(|g\rangle-|gh\rangle)(|g'\rangle+|g'h\rangle)|10\rangle|1\rangle_a\\
&+&(|g\rangle-|gh\rangle)(|g'\rangle-|g'h\rangle)|11\rangle|1\rangle_a
\Big].
\end{eqnarray*}
Let us denote this state by
\begin{eqnarray*}
\frac{1}{4N2^{t(n)}}
\Big[
|h_+\rangle|h_+\rangle|00\rangle|0\rangle_a
+|h_+\rangle|h_-\rangle|01\rangle|1\rangle_a
+|h_-\rangle|h_+\rangle|10\rangle|1\rangle_a
+|h_-\rangle|h_-\rangle|11\rangle|1\rangle_a
\Big],
\end{eqnarray*}
where 
\begin{eqnarray*}
|h_\pm\rangle&\equiv&\sum_{g\in H}\gamma_g(|g\rangle\pm|gh\rangle).
\end{eqnarray*}
The square of the norm of the state, i.e., the postselection probability, is
\begin{eqnarray}
P(p=1)&=&\frac{1}{16N^22^{2t(n)}}
(\langle h_+|h_+\rangle+\langle h_-|h_-\rangle)^2\nonumber\\
&=&\frac{(\sum_{g\in H}\gamma_g^2)^2}{N^22^{2t(n)}}\label{eq2}\\
&=&\frac{N^2(\sum_{g\in H}\gamma_g^2)^2}{N^42^{2t(n)}}\nonumber\\
&=&\frac{N^2(\sum_{g\in H}\frac{\gamma_g^2}{N^2})^2}{2^{2t(n)}}\nonumber\\
&=&\frac{1}{2^{2t(n)-2s(n)}}
\Big(\frac{1}{|H|}+\sum_{g\in H}\epsilon_g^2\Big)^2\label{eq3},
\end{eqnarray}
where we mean $p=1$ if the garbage registers are
projected on to $|+\rangle^{\otimes 2t(n)}$,
and we have used $N=2^{s(n)}$, Eq.~(\ref{eq1}), and the relation
\begin{eqnarray*}
	\langle h_+|h_+\rangle
	+\langle h_-|h_-\rangle
	=4\sum_{g\in H}\gamma_g^2.
\end{eqnarray*}
Therefore, from Eq.~(\ref{eq2}), the normalized state after the postselection
is
\begin{eqnarray*}
\frac{1}{4\sum_{g\in H}\gamma_g^2}
\Big[|h_+\rangle|h_+\rangle|00\rangle|0\rangle_a
+|h_+\rangle|h_-\rangle|01\rangle|1\rangle_a
+|h_-\rangle|h_+\rangle|10\rangle|1\rangle_a
+|h_-\rangle|h_-\rangle|11\rangle|1\rangle_a
\Big].
\end{eqnarray*}
If we project the ancilla qubit onto $|0\rangle_a$,
the (unnormalized) state after the projection is 
\begin{eqnarray*}
&&\frac{1}{4\sum_{g\in H}\gamma_g^2}
|h_+\rangle|h_+\rangle|00\rangle,
\end{eqnarray*}
and therefore, 
\begin{eqnarray*}
	P(o=0|p=1)=
\Big(\frac{\langle h_+|h_+\rangle}{4\sum_{g\in H}\gamma_g^2}\Big)^2.
\end{eqnarray*}
Here, we mean $o=0$ (resp., $o=1$) 
if the ancilla qubit is projected onto $|0\rangle_a$ (resp., $|1\rangle_a$).
If $h\notin H$, 
\begin{eqnarray*}
	\langle h_+|h_+\rangle=2\sum_{g\in H}\gamma_g^2
\end{eqnarray*}
and therefore,
\begin{eqnarray*}
	P(o=0|p=1)&=&\frac{1}{4},\\
	P(o=1|p=1)&=&1-P(o=0|p=1)=\frac{3}{4}.
\end{eqnarray*}
If $h\in H$, on the other hand,
\begin{eqnarray}
P(o=1|p=1)&=&1-P(o=0|p=1)\nonumber\\
&=&1-\Big(\frac{\langle h_+|h_+\rangle}{4\sum_{g\in H}\gamma_g^2}\Big)^2\nonumber\\
&=&1-\Big(\frac{4\sum_{g\in H}\gamma_g^2-\langle h_-|h_-\rangle}{4\sum_{g\in H}\gamma_g^2}\Big)^2\nonumber\\
&=&1-\Big(1-\frac{\langle h_-|h_-\rangle}{4\sum_{g\in H}\gamma_g^2}\Big)^2\nonumber\\
&=&1-\Big(1-\frac{\langle h_-|h_-\rangle}{2\sum_{g\in H}\gamma_g^2}
+\Big(\frac{\langle h_-|h_-\rangle}{4\sum_{g\in H}\gamma_g^2}\Big)^2\Big)\nonumber\\
&\le&\frac{\langle h_-|h_-\rangle}{2\sum_{g\in H}\gamma_g^2}\nonumber\\
&=&\frac{\sum_{g\in H}(\gamma_g-\gamma_{gh^{-1}})^2}{2\sum_{g\in H}\gamma_g^2}\nonumber\\
&\le&\frac{4\epsilon^2|H|}{2\sum_{g\in H}\frac{\gamma_g^2}{N^2}}\nonumber\\
&\le&2\epsilon^2|H|^2\label{eq100}.
\end{eqnarray}
Here, we have used Eqs.~(\ref{eq0}) and (\ref{eq1.1}), and
\begin{eqnarray*}
\langle h_-|h_-\rangle&=&\sum_{g,g'\in H}\gamma_g\gamma_{g'}
	(\langle g|-\langle gh|)
	(|g'\rangle-|g'h\rangle)\\
&=&\sum_{g,g'\in H}\gamma_g\gamma_{g'}
	(\delta_{g,g'}-\delta_{g,g'h}
	-\delta_{gh,g'}+\delta_{g,g'})\\
	&=&\sum_{g\in H}\gamma_g^2
	-\sum_{g\in H}\gamma_g\gamma_{gh^{-1}}
	-\sum_{g'\in H}\gamma_{g'h^{-1}}\gamma_{g'}
	+\sum_{g\in H}\gamma_g^2\\
	&=&\sum_{g\in H}\gamma_g^2
	-\sum_{g\in H}\gamma_g\gamma_{gh^{-1}}
	-\sum_{g\in H}\gamma_{gh^{-1}}\gamma_{g}
	+\sum_{g\in H}\gamma_{gh^{-1}}^2\\
	&=&\sum_{g\in H}(\gamma_g-\gamma_{gh^{-1}})^2.
\end{eqnarray*}


Now we use the result by Fortnow and Rogers~\cite{FR}:
\begin{theorem}
	For any uniform family of polynomial-size quantum circuits,
	there exist $g\in{\rm GapP}$ and a polynomial $q$ such that
	for any $w$, the output probability of the quantum
	circuit on input $w$ is equal to $g(w)/2^{q(|w|)}$.
	(Note that this theorem depends on the gate set.
	In this paper, we consider the Hadamard and Toffoli gates
	as a universal gate set.)
\end{theorem}

From this theorem, there exists a GapP function $g$ and a polynomial $q$
such that
\begin{eqnarray}
	P(o=1,p=1)=\frac{g(w)}{2^{q(n)}},
	\label{GapPrelation}
\end{eqnarray}
where $w$ is an input of the modified GNM.
In the above, we have shown that if $w$ is a yes instance
of the modified GNM, which means $h\notin H$, 
\begin{eqnarray*}
	\frac{3}{4}=P(o=1|p=1)\le 1,
\end{eqnarray*}
which means
\begin{eqnarray*}
	\frac{3}{4}P(p=1)=P(o=1,p=1)\le P(p=1).
\end{eqnarray*}
From Eq.~(\ref{GapPrelation}), it is
\begin{eqnarray*}
	\frac{3}{4}P(p=1)=\frac{g(w)}{2^{q(n)}}\le P(p=1).
\end{eqnarray*}
From Eq.~(\ref{eq3}) and $|H|\le2^n$, this means
\begin{eqnarray*}
\frac{1}{2^{2t(n)-2s(n)}}
\Big(\frac{1}{|H|}+\sum_{g\in H}\epsilon_g^2\Big)^2\frac{3}{4}
=\frac{g(w)}{2^{q(n)}}
\le\frac{1}{2^{2t(n)-2s(n)}}\Big(\frac{1}{|H|}+\sum_{g\in H}\epsilon_g^2\Big)^2
\end{eqnarray*}
$\Leftrightarrow$
\begin{eqnarray*}
\Big(1+|H|\sum_{g\in H}\epsilon_g^2\Big)^2\frac{3}{4}
=\frac{g(w)2^{2t(n)-2s(n)}|H|^2}{2^{q(n)}}
\le\Big(1+|H|\sum_{g\in H}\epsilon_g^2\Big)^2
\end{eqnarray*}
$\Rightarrow$
\begin{eqnarray*}
\frac{3}{4}
\le\frac{g(w)2^{2t(n)-2s(n)}|H|^2}{2^{q(n)}}
\le\Big(1+|H|^2\epsilon^2\Big)^2
\end{eqnarray*}
$\Rightarrow$
\begin{eqnarray*}
\frac{3}{4}
\le\frac{g(w)2^{2t(n)-2s(n)}|H|^2}{2^{q(n)}}
\le(1+2^{2n}\epsilon^2)^2
\end{eqnarray*}
$\Leftrightarrow$
\begin{eqnarray*}
\frac{3}{4(1+2^{2n}\epsilon^2)^2}
\le\frac{g(w)2^{2t(n)-2s(n)}|H|^2}{2^{q(n)}(1+2^{2n}\epsilon^2)^2}
\le1.
\end{eqnarray*}
If we take $\epsilon=2^{-n-3}$ in Theorem~\ref{theorem:Babai},
\begin{eqnarray*}
\frac{2}{3}<\frac{3}{4(1+2^{2n}\epsilon^2)^2}.
\end{eqnarray*}
If we define 
\begin{eqnarray*}
G(w)=g(w)2^{2t(n)-2s(n)}|H|^2,
\end{eqnarray*}
and
\begin{eqnarray*}
F(w)=2^{q(n)}(1+2^{2n}\epsilon^2)^2,
\end{eqnarray*}
we thus obtain
\begin{eqnarray}
	\frac{2}{3}\le \frac{G(w)}{F(w)}\le 1.
	\label{AWPP1}
\end{eqnarray}
By the definition of the modified GNM, 
$|H|\in {\rm FP}\subseteq{\rm GapP}$.
Since GapP functions are closed under multiplications,
$G\in{\rm GapP}$.
Furthermore, since we can assume $q(n)\ge12$ for all~$n$ without loss of generality,
we have $F\in{\rm FP}$ for our choice of $\epsilon$.

On the other hand, if $w$ is a no instance of the
modified GNM, which means $h\in H$,
we obtain from Eqs.~(\ref{GapPrelation}) and (\ref{eq100}) that
\begin{eqnarray*}
0\le\frac{g(w)}{2^{q(n)}}\le 2\epsilon^2|H|^2P(p=1)
\end{eqnarray*}
$\Leftrightarrow$
\begin{eqnarray*}
0\le\frac{g(w)}{2^{q(n)}}
\le2\epsilon^2|H|^2\frac{1}{2^{2t(n)-2s(n)}}
\Big(\frac{1}{|H|}+\sum_{g\in H}\epsilon_g^2\Big)^2
\end{eqnarray*}
$\Rightarrow$
\begin{eqnarray*}
0\le\frac{g(w)2^{2t(n)-2s(n)}|H|^2}{2^{q(n)}}
\le2\epsilon^2|H|^2
\Big(1+\epsilon^2|H|^2\Big)^2
\end{eqnarray*}
$\Leftrightarrow$
\begin{eqnarray*}
0\le\frac{g(w)2^{2t(n)-2s(n)}|H|^2}{2^{q(n)}(1+2^{2n}\epsilon^2)^2}
\le2\epsilon^2|H|^2.
\end{eqnarray*}
Since $\epsilon=2^{-n-3}$,
\begin{eqnarray*}
	2\epsilon^2|H|^2\le2^{-5}\le\frac{1}{3}.
\end{eqnarray*}
We thus obtain
\begin{eqnarray}
	0\le\frac{G(w)}{F(w)}\le \frac{1}{3}.
	\label{AWPP2}
\end{eqnarray}
Since Eqs.~(\ref{AWPP1}) and (\ref{AWPP2}) satisfy
the definition of AWPP, we conclude that
the modified GNM is in AWPP.

\acknowledgements
TM is supported by the JSPS Grant-in-Aid for Young Scientists (B)
No.26730003 and the MEXT JSPS Grant-in-Aid for Scientific Research on
Innovative Areas No.15H00850.
HN is supported by the JSPS Grant-in-Aid for Scientific Research (A)
Nos.23246071, 24240001, 26247016, and (C) No.25330012, and
the MEXT JSPS Grant-in-Aid for Scientific Research on Innovative Areas
No.24106009. FLG is supported by the JSPS Grant-in-Aid for Young 
Scientists~(B) No.~24700005, the JSPS Grant-in-Aid for Scientific 
Research~(A) No.~24240001, and the MEXT JSPS Grant-in-Aid for Scientific 
Research on Innovative Areas No.~24106009.

\end{document}